\documentclass[a4paper]{article}
\usepackage{RR}
\usepackage{hyperref}
\RRdate{Mars 2009}

\usepackage{graphicx,amsmath,color,float}
\usepackage{amsfonts}
\usepackage{amssymb}

\RRauthor{J\'er\^ome Jaffr\'e  \thanks{INRIA, BP 105, 78153 Le
Chesnay Cedex, France (Jerome.Jaffre@inria.fr)} \and
Amel Sboui \thanks{Andra, 1-7, rue Jean Monnet, 92298 Ch\^atenay-Malabry cedex, France,
\emph{Present address:} 
LaMSID - UMR EDF/CNRS 2832,
EDF R\&D,
1, Avenue du g\'en\'eral de Gaulle, 92141 Clamart Cedex, France (Amel.Sboui@gmail.com)}}
\authorhead{J\'er\^ome Jaffr\'e and Amel Sboui}
\RRetitle{Henry' law and gas phase disappearance\thanks{This work was partially supported by GdR MoMaS (PACEN/CNRS, ANDRA, BRGM, CEA, EDF, IRSN).}}
\RRtitle{La loi de Henry et la disparition de la phase gazeuse}
\titlehead{Henry' law and gas phase disappearance}
\newcommand{\R}{{\mathbb R}}

\numberwithin{equation}{section}
\numberwithin{figure}{section}
\newcommand{\red}{\textcolor{black}}
\newcommand{\blue}{\textcolor{black}}

 \newcommand{\p}{\partial}
 
\renewcommand{\div}{\mbox{ div}}
 
 \newcommand{\grad}{{\nabla}}

\RRabstract{
For a two-phase (liquid-gas) two-component (water-hydrogen) system we discuss the formulation of the possible dissolution of hydrogen in the liquid phase. The problem is formulated as a set of nonlinear partial differential equations with complementary constraints and we show how Henry's law fits in a phase diagram.}
\RRkeyword{Porous media, two-phase flow, dissolution, nuclear waste underground storage}
\RRresume{Pour un syst\`eme diphasique (liquide-gaz) \`a deux composants (eau-hydrog\`ene) on consid\`ere la formulation de la dissolution \'eventuelle de l'hydrog\`ene dans la phase aqueuse. On montre comment la loi de Henry entre dans un diagramme de phase. Le probl\`eme est alors formul\'e comme un ensemble d'\'equations aux d\'eriv\'ees partielles non-lin\'eaires avec des contraintes de comp[l\'ementarit\'e.
}
\RRmotcle{Milieu poreux, \'ecoulement diphasique, dissolution, stockage profond de d\'echets nucl\'eaires}
\RRprojet{Estime}
\RRtheme{\THNum}
\RCParis

\begin{document}
\RRNo{6891}
\makeRR
\section{Introduction}
Recently the Couplex-Gas benchmark (http://www.gdrmomas.org/ex\_qualifications.html) was proposed by Andra and MoMAS in order to improve the simulation of  the migration of hydrogen produced by the corrosion of nuclear waste packages in an underground storage. This is a system of two-phase (liquid-gas) flow with two components (hydrogen-water). This is similar to a black-oil model but the benchmark was proposing to use in Henry's law to model the dissolution of hydrogen.  This benchmark generated some interest and engineers encountered difficulties in handling the appearance and disappearance of the phases. Recently two other papers 
\cite{BourgeatJurakSmai}, \cite{AbadpourPanfilov} 
adressed the formulation of this problem. 

In this paper, for the sake of the simplicity of exposition, we will assume that only hydrogen is in the gas phase (water does not vaporize). We discuss how to extend Henry's law to the case where the gas phase disappears. This is done by showing how Henry's law fits in a phase diagram. The resulting formulation is a set of partial differential equations with  complementarity constraints. A similar formulation is also presented in \cite{BourgeatJurakSmai} and it also can be viewed as a variation of the formulation of the black oil model presented in \cite{chajaf86}.

\section{Mathematical formulation}
We first describe the equations and some simplifying hypotheses. Since we consider a problem where the gas phase can disappear while the liquid phase cannot disappear we will consider a formulation using the saturation and the pressure of the liquid phase as the two main unknowns. 
\subsection{Fluid phases}
$\ell$ and $g$ are the respective indices for the liquid phase and the gas phase.
Darcy's law reads
\begin{equation}
{\bf q}_i = -K(x) k_i(s_i) (\grad p_i - \rho_i g \grad z), \quad i=\ell, g,
\label{Darcy}
\end{equation}
where $K$ is the absolute permeability.  Assuming that the phases occupy the whole pore space, the phase saturations $s_i, i=\ell,g$ satisfy
\[ 0 \leq s_i \leq 1, \quad s_\ell+s_g=1. \]
The mobilities are denoted by $k_i(s_i)= \dfrac{{k_r}_i(s_i)}{\mu_i}$ with $k_{ri}$ the relative permeability, and $\mu_i$ the viscosity of the phase $i=\ell, g$, assumed to be constant. The mobility $k_i$ is an increasing function of $s_i$ such that $k_i(0)=0, i=\ell, g$.

The phase pressures are related through the capillary pressure law
\begin{equation}
p_c(s_\ell) = p_g -p_\ell  \geq 0.
\label{pc}
\end{equation}
assuming that the gas phase is the nonwetting phase. The capillary pressure is a decreasing function
of the saturation $s_\ell$ such that $p_c(1) = 0$.

In the following we will choose $s_\ell, p_\ell$ as the main variables since we assume that the liquid phase cannot disappear for the problem under consideration.

\subsection{Fluid components}
We consider two components, water and hydrogen denoted by the indices $j=w, h$. The mass density of the phase $i$ is 
\[
\rho_i = \rho_w^i + \rho_h^i, \quad  i=\ell, g,
\]
where $\rho_w^i$ and $\rho_h^i$ are the mass concentrations of water and of hydrogen in the phase $i, i=\ell,g$.
The mass fractions are
\[
\chi_h^i = \frac{\rho_h^i}{\rho_i},\quad \chi_w^i = \frac{\rho_w^i}{\rho_i}, \quad  i=\ell,g.
\]
Obviously we have
\[
\chi_w^i + \chi_h^i = 1,\quad  i=\ell,g .
\]
We assume that the liquid phase may contain both components, while the gas phase contains only hydrogen, that is the water does not vaporize.  In this situation we have
\[
\rho_w^g=0, \quad \rho_g = \rho_h^g, \quad \chi_h^g = \frac{\rho_h^g}{\rho_g} =1 ,\quad \chi_w^g =0.
\]
A third main unknown will be $\chi_h^\ell$.

\subsection{Conservation of mass}
We introduce the molecular diffusion fluxes $j_i^\ell, i=w,h$ for the diffusion of component $i$ in the liquid phase. They must satisfy $j_h^\ell + j_w^\ell =0$ and we have
\begin{equation}
j_h^\ell = -\phi s_\ell \rho_\ell D_h^\ell\grad\chi_h^\ell. \end{equation}
The coefficients $D_h^\ell$  is a molecular diffusion coefficient.

Conservation of  mass applied to each component, water and hydrogen, gives 
\begin{equation} \begin{array}{l}
\phi \dfrac{\p}{\p t} (s_\ell \rho_\ell \chi_w^l) + \div(\rho_\ell \chi_w^l{\bf q}_\ell + j_w^\ell) = Q_w, \\[0.3cm]
\phi \dfrac{\p}{\p t} (s_\ell \rho_\ell \chi_h^\ell +s_g\rho_g) + \div(\rho_\ell \chi_h^\ell {\bf q}_\ell + \rho_g {\bf q}_g + j_h^\ell) = Q_h.
\label{cons}
\end{array} \end{equation}
We assume also that the gas is slightly compressible, that is $\rho_g = C_g p_g$ with $C_g$ the compressibility constant, and that the liquid phase is incompressible, that is $\rho^\ell$ is constant.

\section{Phase equilibrium}
\subsection{Phase diagram}
\begin{figure}[htbp]
\begin{center}
\includegraphics[width=6cm]{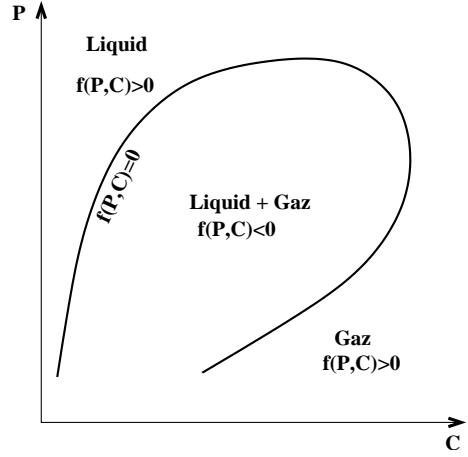}
\caption{A standard phase diagram.}
\label{phasediagram}
\end{center}
\end{figure}

A phase diagram is a figure like that shown in  Figure \ref{phasediagram}
where a curve $f(P,C)=0$ separates two zones, a first zone defined by $f(P,C)>0$ where the systeme is monophasic  -- liquid or gas -- and a second zone defined by $f(P,C)<0$ where the system is diphasic 
 -- liquid + gas. It gives the repartition of two components, like hydrogen and water, into the two phases, liquid and gas. 

For  the example we are concerned with, that is the eventual disappearance of the gas phase, we are interested in the left part of the phase diagram. There we have either $s_l =1$ and $f(P,C) \geq 0$ in the liquid zone, or $s_l < 1$ and $f(P,C) \leq 0$ in the diphasic zone. This can be written in terms of the following set of complementarity constraints
\[ (1-s_l)f(P,C) = 0, 1-s_l \geq 0, f(P,C) \geq 0. \]

The variable $C$ is the total concentration of hydrogen, that is in the diphasic zone $C=\rho_h^\ell + \rho_g$ (since we assumed that only hydrogen can be present in the gas phase), in the liquid zone liquide  $C=\rho_h^\ell$ and in the gas zone $C=\rho_g$ since we assumed that the water cannot vaporize. 

The variable $P$ is called the  "gas pressure",  $P=p_g$ in the gas and diphasic zone. To extend this definition to the liquid zone we express $p_g$ in terms of the capillary  pressure, $p_g =  p_\ell+p_c(s_\ell)$, and define 
$P=p_\ell+p_c(s_\ell)$. Thus in the liquid zone we have $s_\ell=1, p_c(s_\ell)=0$ et $P=p_\ell$.
The definitions of the variables $P$ and $C$ are shown in Table~\ref{tvar}.
\begin{table}[htdp]
\begin{center}
\begin{tabular}{|c|c|c|c|c|c|}
\hline
  & liquid & liquid + gas & gas \\ \hline
  & $s_\ell =1$ & $0< s_\ell < 1$ & $s_\ell =0$\\
 \hline
 $P=$ & $p_\ell$ & $p_g = p_\ell+p_c(s_\ell)$& $p_g$ \\
 \hline
 $C=$ & $\rho_h^\ell$ & $\rho_h^\ell + \rho_g$&$\rho_g$ \\
 \hline
 $f(P,C)$ & $\geq 0$ & $\leq 0$&$\geq 0$ \\
 \hline
\end{tabular}
\end{center}
\caption{Definition of the variables $P$ et $C$ in the phase diagram.}
\label{tvar}
\end{table}

\subsection{Extension of Henry's law}
In the following we focus on the interface between the liquid zone and the diphasic zone.
In the presence of the gas phase Henry's law reads
\begin{equation} H p_g  = \rho_h^\ell \label{Henry} \end{equation}
with $H$ the Henry constant.
We can integrate this relation into a phase diagram formulation which includes the situation with no gas phase. This formulation includes a set of complimentarity constraints.

\begin{figure}[htbp]
\begin{center}
\includegraphics[width=6cm]{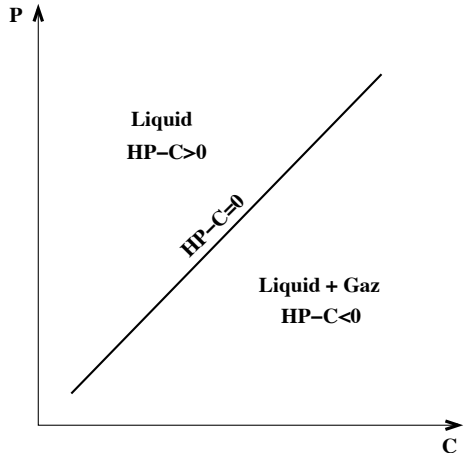}
\caption{Part of the phase diagram separating the liquid and diphasic zones when Henry's law is used.}
\label{henrydiagram}
\end{center}
\end{figure}

Henry's law is defined only in the diphasic zone. In this zone we have
$HP - C = Hp_g - (\rho_h^\ell + \rho_g) = - \rho_g < 0 $,
and when we move to the curve separating the  liquid and diphasic zones we have
$HP - C = \rho_g \rightarrow 0$. Above this curve is the liquid zone 
$HP - C = Hp_\ell - \rho^\ell_h > 0$. Thus using Henry's law is actually  replacing the part of the curve $f(P,C)=0$ which separates the liquid and diphasic zones  by the straight line $HP-C=0$. This is shown in Figure \ref{henrydiagram} where only the left part of the phase diagram is shown with the line separating the liquid zone and the diphasic zone.

Indeed, either the gas phase exists, $1-s_\ell > 0$, Henry's law applies and $H(p_\ell + p_c(s_\ell)) - \rho_h^\ell =0$,  or the gas phase does not
exist, $s_\ell = 1, p_c(s_\ell) = 0$ and $Hp_\ell  - \rho_h^\ell > 0$ which says that for a given pressure 
$p_\ell$ the concentration is too small for the hydrogen component to be partly gaseous, or conversely
for a given concentration $\rho_h^\ell$ the pressure $p_\ell$ is too large for the hydrogen component to be partly gaseous. These cases can be written as the complementarity constraints
\begin{equation}  (1-s_\ell)\Bigl(H(p_\ell + p_c(s_\ell)) - \rho_h^\ell\bigr) =0,  \quad  1-s_\ell \geq 0, 
\quad H(p_\ell + p_c(s_\ell)) - \rho_h^\ell \geq 0. 
\label{comp}
\end{equation}

\section{Formulation with complementarity constraints}

Finally we end up with a system of nonlinear partial equations -- conservation equations and Darcy laws -- with nonlinear complementarity constraints describing  the transfer of hydrogen between the two phases, the unknowns being  $s_\ell, p_\ell, \chi_w^\ell$: 
\begin{equation} \begin{array}{ll}
\phi \dfrac{\p}{\p t} \bigl(\red{s_\ell} \rho_\ell (1-\red{\chi_h^\ell})\bigr) + 
\div\bigl(\rho_\ell (1-\red{\chi_h^\ell}) {\bf q}_\ell -j_h^\ell \bigr) = Q_w \\[0.3cm]
\phi \dfrac{\p}{\p t} \bigl(\red{s_\ell} \rho_\ell\red{ \chi_h^\ell} + 
(1-\red{s_\ell})C_g(\red{p_\ell} +p_c(\red{s_\ell}))\bigr) + \\
\hspace*{4cm} \div \bigl(\rho_\ell \red{\chi_h^\ell} {\bf q}_\ell + 
C_g(\red{p_\ell} +p_c(\red{s_\ell})) {\bf q}_g + j_h^\ell \bigr) = Q_h\\[0.3cm]
{\bf q}_\ell = -K(x) k_\ell(\red{s_\ell}) (\grad \red{p_\ell} - \rho_\ell g \grad z)\\[0.3cm]
{\bf q}_g = -K(x) k_g(1-\red{s_\ell}) (\grad (\red{p_\ell} + p_c(\red{s_\ell})) 
- C_g(\red{p_\ell} + p_c(\red{s_\ell})) g \grad z)\\[0.3cm]
j_h^\ell = -\phi \red{s_\ell} \rho_\ell D_h^\ell\grad\red{\chi_h^\ell}\\[0.3cm]
(1-\red{s_\ell})\bigl(H(\red{p_\ell} + p_c(\red{s_\ell})) - \rho_\ell \red{\chi_h^\ell}\bigr) =0,  \quad  1-\red{s_\ell} \geq 0, \quad
 H(\red{p_\ell} + p_c(\red{s_\ell})) - \rho_\ell \red{\chi_h^\ell} \geq 0.
\end{array} \label{system} \end{equation}
This formulation is valid whether the gas phase exists or not. A similar system was presented in \cite {BourgeatJurakSmai}. We note that when the gas phase disappears, i.e. $s_\ell =1$, the second equation in (\ref{system}) becomes a standard transport equation of hydrogen as a solute in the liquid phase.

Assume that we discretize problem (\ref{system}) with cell centered finite volumes and we denote by
$N$, the number of degrees of freedom for $\red{s_\ell}, \red{p_\ell}, \red{\chi_h^\ell}$ which is equal to the number of cells. We 
introduce
\[ \begin{array}{l}
\blue{x} \in \R^{3N}, \mbox{ the vector of unknowns for } \red{s_\ell}, \red{p_\ell}, \red{\chi_h^\ell},\\
\blue{{\cal H}}: \R^{3N} \rightarrow \R^{2N}, \mbox{ the vector of the discretized component conservation equations,}\\
\blue{{\cal F}}: \R^{3N} \rightarrow \R^{N}, \mbox{ the discrete vector for } 1-\red{s_\ell},\\
\blue{{\cal G}}: \R^{3N} \rightarrow \R^{N}, \mbox{ the discrete vector for } H(\red{p_\ell} + p_c(\red{s_\ell})) - \rho_\ell \red{\chi_h^\ell}.
\end{array} \]
Then the problem can be written in compact form
\[ \begin{array}{l}
\blue{{\cal H}(x) =0,}\\[0.2cm]
\blue{{\cal F}(x)^{\,\top} {\cal G}(x) =0,\quad {\cal F}(x) \geq 0, \quad {\cal G}(x) \geq 0.}
\end{array} \]
Building a solver to solve properly such a problem is our current task. This task is undertaken in the field of numerical optimization. For a survey on complementarity problems see [4].

There are clearly alternative ways to solve this problem without using such a solver. See for recent examples \cite{BourgeatJurakSmai}, \cite{AbadpourPanfilov}. However nonlinear problems with complementary constraints occur in many circumstances as for example in a black oil model \cite{chajaf86} or in problems with dissolution-precipitation. In  \cite{Krautle} this latter problem was solved using a semi-smooth Newton method inspired by \cite{Kanzow}. Therefore we believe that it is worthwhile to build a general purpose appropriate solver for nonlinear problems with complementarity constraints, and the problem of a liquid-gas system with dissolution of hydrogen is a very good test problem.

\section{Conclusion}
We showed how to extend Henry's law to the case where the gas phase disappears by plugging it into a phase diagram formulation. This results into the formulation of a two-phase (liquid-gas) two-component (water-hydrogen) system with possible dissolution of hydrogen as a system of nonlinear partial differential equations with a set of nonlinear complementarity constraints.

\bibliographystyle{plain} 

\begin{thebibliography}{}
\bibitem{AbadpourPanfilov}
A. Abadpour and M. Panfilov,
Method of negative saturations for multiple compositional flow with oversaturated zones.
{\em Transport in Porous Media} (2009).
\bibitem{BourgeatJurakSmai}
A. Bourgeat, M. Jurak and F. Sma\"{\i},
Two phase partially miscible flow and transport modeling in
porous media; application to gas migration in a nuclear waste repository.
{\em Computational Geosciences} 13 (2009), pp. 29-42..
\bibitem{chajaf86}
G. Chavent and J. Jaffr\'e,
{\em Mathematical Models and Finite Elements for Reservoir Simulation},
Studies in Mathematics and its Applications 17 (North Holland, Amsterdam, 1986).
\bibitem{FacchineiPang}
F. Facchinei and J. S. Pang,
{\em Finite Dimensional Variational Inequalities and Complementarity Problems}
(Springer, 2003).
\bibitem{Kanzow}
C. Kanzow, Inexact semismooth {Newton} methods for large-scale complementarity problems. {\em Optimization Methods and Software} 19 (2004), 
pp. 309-325.
\bibitem{Krautle}
S. Kra\"{u}tle, {\em General Multi-Species Reactive Transport Problems in Porous Media: 
Efficient Numerical Approaches and Existence of Global Solutions.}
University of {Erlangen-Nuremberg}, Department of Mathematics 
(March 2008).
\end{thebibliography}

\end{document}